\documentclass[twocolumn,preprintnumbers,amsmath,amssymb]{revtex4}

\usepackage{graphicx}
\usepackage{dcolumn}
\usepackage{bm}
\begin{document}


\title{Effect of Coulomb scattering on graphene conductivity}
\author{V. Vyurkov$^{{\rm 1}}$}
\email{vyurkov@ftian.ru}
\author{V. Ryzhii$^{{\rm 2}}$}%
\email{v-ryzhii@u-aizu.ac.jp}
\affiliation{%
$^{{\rm 1}}$Institute of Physics and Technology RAS, Moscow 117218, Russia\\
$^{{\rm 2}}$University of Aizu, Aizu-Wakamatsu 965-8580, Japan,
and Japan Science and Technology Agency, CREST, Tokyo 107-0075,
Japan }
\date{July 24, 2008}

\begin{abstract}

The effect of Coulomb scattering on graphene conductivity in field effect transistor structures is discussed. Inter-particle scattering (electron-electron, hole-hole, and electron-hole) and scattering on charged 
defects are taken into account in a wide range of gate voltages. 
It is shown that an intrinsic conductivity of graphene (purely
ambipolar system where both electron and hole densities exactly
coincide) is defined by strong electron-hole scattering. It has a
universal value independent of temperature. We give an explicit
derivation based on scaling theory. When there is even a small
discrepancy in electron and hole densities caused by applied gate
voltage the conductivity is determined by both strong electron-hole scattering and weak external scattering: on defects or phonons. We suggest that a density of charged defects (occupancy of defects) depends on Fermi
energy to explain a sub-linear dependence of conductivity on a
fairly high gate voltage observed in experiments. We also eliminate contradictions between experimental data obtained in deposited and suspended graphene structures regarding graphene conductivity.  
\end{abstract}
\maketitle


Different heterostructures on the basis of {\it Graphene}, i.e., a
monolayer of carbon atoms forming a dense honeycomb
two-dimensional crystal structure are considered as promising
candidates for future micro- and nanoelectronics.
(see~\cite{1,2,3,4,5} and references therein). The features of the
electron and hole energy spectra in graphene provide the
exceptional properties of graphene-based heterostructures and
devices, in particular, field-effect transistors~\cite{4,6,7,8}.
Graphene band structure peculiarities have a substantial impact on 
Coulomb scattering: inter-particle scattering 
(electron-electron, hole-hole, and electron-hole) and scattering on charged 
defects. Here we intend to elucidate the effect of that kind of scattering on
graphene conductivity taking into account either deposited and suspended graphene structures. 

Experimental studies of graphene have revealed a quite amazing
feature of its conductivity. There is almost no
dependence of the minimum conductivity of graphene $\sigma _{min}$ on
temperature (between 0.3K and 300K)in deposited graphene structures~\cite{3} (Fig.1, upper panel). This is really
intriguing because at the same time the carrier density varies in
this temperature interval by six orders of magnitude. The minimum conductivity (or intrinsic conductivity) arises at zero gate voltage
in gated graphene structures. Hereafter, as usual, we reference
the zero gate voltage ($V_{G} = 0)$ just to this point. The
conductivity is approximately equal to $\sigma_{min} \simeq
(6~k\Omega)^{-1}$. There appears a natural temptation to bind this value
to two conductance quanta~\cite{3,9}. The conductance
quantum for spin-unpolarized electrons equals $G_0 = 2e^2/h =
(12k\Omega)^{-1}$, where $e$ is the electron elementary charge and
$h$ is the Planck constant. 

\begin{figure}
\centerline{\includegraphics[width=70mm]{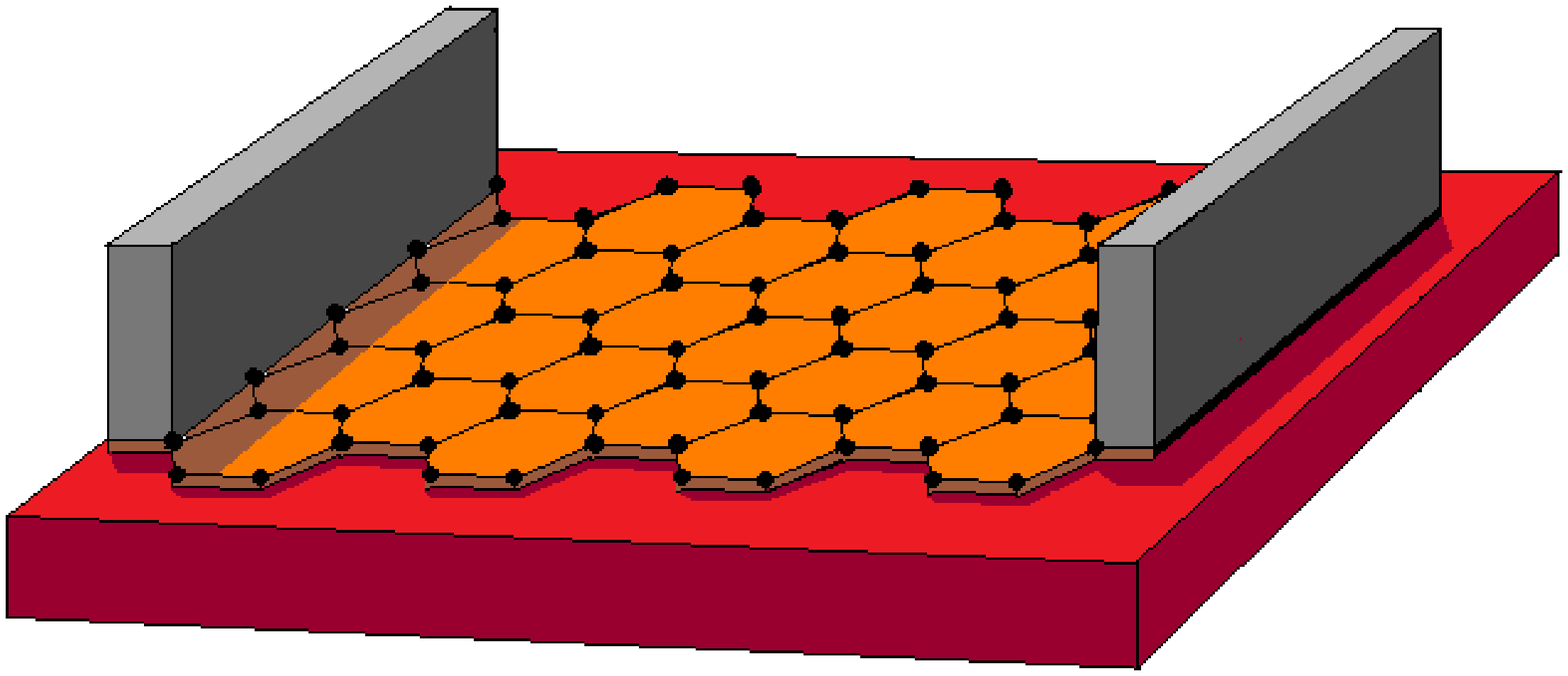}}
\end{figure}
\begin{figure}
\centerline{\includegraphics[width=70mm]{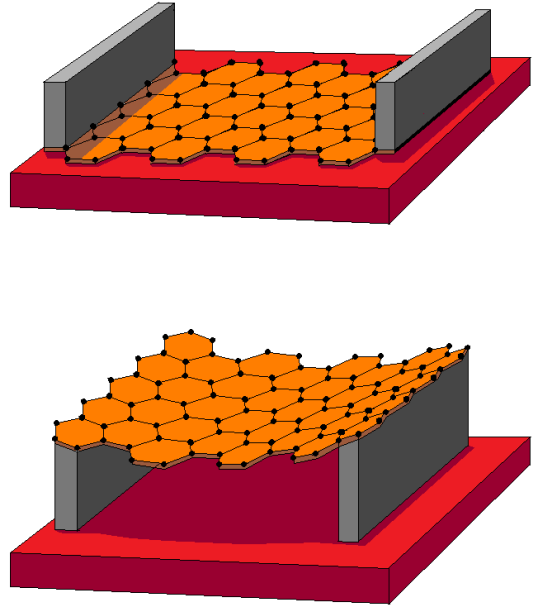}}

\caption{Schematic view of graphene structures 
with two side contacts and a wafer as a back gate: deposited graphene 
(upper panel) and suspended graphene(lower panel)}.
\label{fig:fig1}
\end{figure}

The minimum conductivity of graphene corresponds to an
utterly ambipolar system when electron density $\Sigma _{e}$ and
hole density $\Sigma _{h}$ coincide: $\Sigma _{e}=\Sigma_{h}$. In
the situation, the electron-hole scattering is the strongest
scattering mechanism. We
propose below a derivation
based on a scaling theory approach (see
also~\cite{10}).

The  conductivity, $\sigma _{0}$, of an intrinsic graphene layer
determined by the
 Coulomb scattering of
electrons and holes can  be  expressed as follows:

\begin{equation}
\label{eq1} \sigma _{0} = \frac{k _{eff}^{2}}{e^2}f_0,
\end{equation}
where $k _{eff}$ is the effective permittivity and $f_0$ is a so
far unknown function of the parameters of the graphene
electron-hole system and, generally speaking, of the temperature.
The proper dependence (1) on the elementary charge $e$ and the
effective permittivity $k _{eff}$ for the Coulomb interaction in
the Born approximation  is factorized. As it was recently
theoretically demonstrated~\cite{11,12,13,14}, the screening in a
graphene layer, turns out to be determined by a weak function of
the dimensionless parameter
\begin{equation}
\label{eq2} \alpha = \frac{e^2}{k _{eff}\, \hbar\, v_{F}},
\end{equation}
where $\hbar$ is the reduced Planck constant.
In reality, the screening by highly conducting gates might be
substantial were they placed fairly close to a graphene sheet.

The factor $f_0$ in~(1) can solely depend on intrinsic graphene
parameters and the Plank constant.
 In the case when electron and hole
densities coincide $\Sigma _{e}=\Sigma _{h}=\Sigma_{T}$,
 the graphene electron-hole system  is fully characterized by
the only parameter besides Fermi velocity $v_{F}$. It could be the
mean momentum $p_T$, mean wave vector $k_T = p_T/\hbar$,
temperature $T \simeq \hbar v_{F} k_T/k$ (where $k$ is the
Boltzmann constant), or density $\Sigma _{T} \simeq k_{T}^{2} $.
All those parameters are directly bound to each other. Therefore,
the only possible combination looks like

\begin{equation}
\label{eq3} \sigma _{0} = \frac{\kappa _{eff}^2 hv_F^2}{e^2} =
\frac{2\kappa _{eff}^2 v_F^2}{G_0}.
\end{equation}

Here we have neglected a numerical factor of the order of unity,
 which could be
derived via thorough consideration of electron-hole
scattering~\cite{15} or determined from experimental data(see, for
instance,~\cite{3}).

The only temperature-dependent parameter $k_T$ describing the
graphene electron-hole system can not be included. Therefore, the
graphene conductance appears to be independent of the temperature.
Moreover, it is readily seen that the graphene conductivity turned
out to be inversely proportional to the conductance quantum $G_0 =
2e^2/h =(13k\Omega)^{-1}$, in spite of the previous
expectations~\cite{3}.

It is expedient to estimate the value of $\sigma_0 $ for the
typical graphene structures. When a graphene sheet is sandwiched
between two dielectrics with permittivity $k_1$ and $k_2$, the
effective permittivity is

\begin{equation}
\label{eq4} \kappa _{eff} = {\frac{{\kappa _{1} + \kappa _{2}}}
{{2}}}.
\end{equation}

Therefore, in our opinion, the crucial experiment to verify the
above speculations lies in changing the permittivity of the
adjacent dielectrics. In most of so far made experiments a
graphene layer was deposited on a SiO$_2$ surface. The effective
permittivity in this case is equal to $k_{eff} \simeq 2.5$. The
velocity $v_F = 10^8$~cm/s corresponds to the conductance equal to
$(3k\Omega)^{-1}$. Substituting all that into (3),
 one obtains $\sigma_0 \simeq (4k\Omega)$.
This is a very good agreement with the experimental value of
$(6k\Omega)^{-1}$~\cite{3} provided a numerical coefficient in
(3) was ignored.

All said above is related to an intrinsic graphene conductivity
when both the electron and hole densities are exactly equal to
each other, i.e., $\Sigma_{e} = \Sigma _{h }= \Sigma _T$, This can
also occur in  graphene-based structures with a highly conducting
gate similar to field-effect transistors. In such a case, the
graphene conductivity is determined, in fact, by the friction
between the  electron and hole subsystems. It appears to be
 interesting to apprehend what occurs beyond this equality which can
be broken by an applied gate voltage. At a moderate gate voltage
there is a rather small difference  in the electron and hole
densities: $|\Sigma_ e - \Sigma _h| \ll \Sigma_e,\,\Sigma _h$,
 so that the electron-hole scattering mechanism
is still the strongest one. However, we argue that in this case,
 the behavior of the
conductivity is governed by both the electron-hole as well as
other scattering processes, for instance, those on phonons and
charged defects. In contarast  to an intrinsic graphene, the
influence of the electron-hole scattering on the conductivity
becomes intermediate. Due to this scattering processes,
 the majority carriers drag the minority
ones. In this case the electron-hole system
 could be regarded as a unified
fluid with a total density equal to $\Sigma = \Sigma _e + \Sigma
_h$ and a renormalized elementary charge of composing particles
given by the following equation:

\begin{equation}
\label{eq5} {\tilde e} = e\frac{\Sigma _{h} - \Sigma _{e}}{\Sigma
_{h} + \Sigma_{e}}.
\end{equation}

Taking into account (5), the conventional formula for the
 conductivity of a two-dimensional electron gas
 can be adopted to the  conductivity of graphene:

\begin{equation}
\label{eq6} \sigma _{1} = \frac{{\tilde e}^2\Sigma} {\nu _{ex} M}
= \frac{e^2}{\nu _{ex} M}\frac{(\Sigma _{e} - \Sigma _{h})^2}
{(\Sigma _{e} + \Sigma _{h} )},
\end{equation}
where $\nu _{ex} $ is the rate of scattering  on phonons and
defects, $M$ is the hydrodynamic mass of particles arising in the
hydrodynamic equations for graphene electron-hole system.
 The latter can be roughly estimated as
\begin{equation}
\label{eq7} M \simeq \frac{\hbar k_T}{v_F} \simeq \frac{\hbar
\sqrt{\Sigma _T}}{v_F}.
\end{equation}

\begin{figure}
\centerline{\includegraphics[width=80mm]{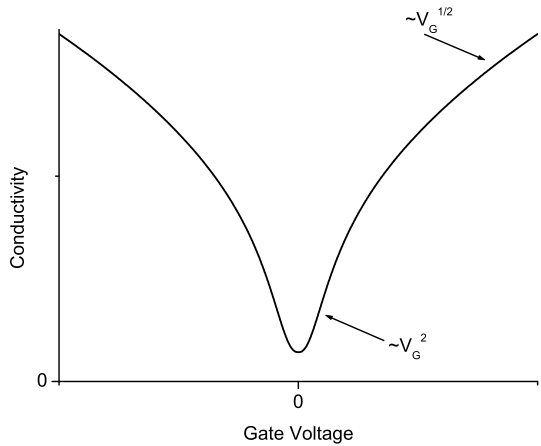}}

\caption{Graphene conductivity vs. gate voltage.}
\label{fig:fig2}
\end{figure}

The conductivity (6) implies a quadratic dependence on gate
voltage $\sigma_{1} \propto V_G^2$ (Fig. 2, low voltage) as a sheet charge $\Sigma_e -
\Sigma_h \propto V_G $. A crossover from intrinsic graphene
conductivity $\sigma_{0}$ given by formula (3) to that given by
formula (5) can be very fast because external scattering $\nu
_{ex}$ is much weaker compared to electron-hole scattering $\nu
_{eh}$ which could be easily extracted from the expression (3). If
among external scattering mechanisms phonon scattering dominates
there appears a definite dependence of conductivity on
temperature. Obviously, the phonon scattering can predominate and
determine conductivity of perfect graphene structures~\cite{16}.
 Then, according to estimations, the electron and hole
mobility can achieve values above 200~000 cm$^2$/V\,s instead of
nowadays 5~000 - 15 000 cm$^{{\rm 2}}$/ V s at room temperature
for deposited graphene structures (Fig. 1, upper panel)~\cite{17}.

Recently a conductivity of suspended graphene (Fig. 1, lower panel) was
experimentally investigated [18]. This is just a structure where
phone scattering can prevail over defect scattering. For the first
time the authors observed temperature dependent minimum graphene
conductivity. However, we attribute this unusual behavior to the
fact that suspended graphene was bent and, therefore,
inhomogeneous with respect to charge distribution. The authors have
admitted this possibility. In other words, there were regions obeying the relation
(3) and that obeying the relation (5) simultaneously. Very high
mobility (100 cm$^2$/V\,s) observed close to the point of
intrinsic graphene when electron and hole densities almost
coincide, really, confirms that phonon scattering is dominant with
respect to that on defects in the case. Nevertheless, below we
demonstrate that at high voltage a proportion of phone scattering
rate to defect scattering rate can become reciprocal.

At fairly high gate voltages when $\Sigma _e \gg \Sigma _h$ or
vice versa, the graphene electron-hole system
 is virtually unipolar and the electron-hole
scattering processes are  not essential. In that case, the
conductivity of the graphene structures fabricated so far is
likely governed by the scattering on charged defects~
\cite{3,14,16},  as there is no temperature dependence of mobility
evidenced by experiments. Nevertheless, theoretical attempts 
to investigate an interplay of phonon scattering and Coulomb 
scattering among particles are made \cite{19}.

The conductivity $\sigma _2 $ limited by the scattering on
charged defects can be obtained almost similarly as that for the
case of the dominant electron-hole scattering. Following~(1), the
conductivity in this situation can be presented as

\begin{equation}
\label{eq8} \sigma _{2} = \frac{k _{eff}^2}{e^2\Sigma _i}f_2,
\end{equation}
where $\Sigma_i$ is the density of charged defects.
If the thermal energy $kT$ is much smaller than the Fermi energy
$\varepsilon _{F}$, the factor $f_{2} $ in (8) can depend only on
the majority carrier density $\Sigma \simeq \Sigma_e$ (or $\Sigma
\simeq \Sigma_h$), the Fermi velocity $v_{F} $, and the Plank
constant $h$, so that

\begin{equation}
\label{eq9} \sigma_{2} = \frac{k_{eff}^2 hv_F^2
\Sigma}{e^2\Sigma_i}.
\end{equation}
For an invariable density of defects, (9) results in a linear
dependence of the conductivity on the gate voltage. On the
contrary, it seems plausible to suggest that the charged defect
concentration can depend on position of the Fermi level in
graphene, i.e., on the Fermi energy $\varepsilon _{F} $ (Fig.3). 

\begin{figure}
\centerline{\includegraphics[width=80mm]{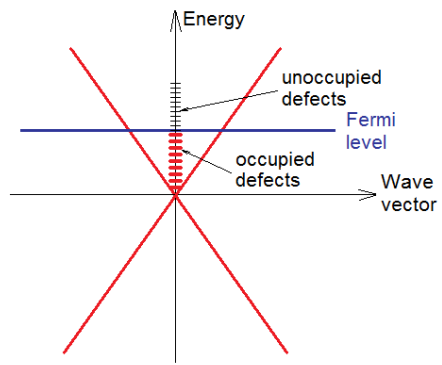}}

\caption{Charged defect states under the Fermi level.}
\label{fig:fig3}
\end{figure}

For
instance, such defects may arise at the graphene-SiO$_2$
interface. If the number of defects per unit energy and area is
constant and equals $n_{i}$, their sheet density is
\begin{equation}
\label{eq10} \Sigma _{i} = n_{i} \varepsilon _{F} \simeq n_{i}
v_{F} \hbar \sqrt {\Sigma}.
\end{equation}
After substituting (10) into (9),  one can arrive at a square-root
dependence of the conductivity on the gate voltage $\sigma_2
\propto \sqrt{V_G}$ (Fig.2, high voltage). The experimental data show that the observed
voltage dependences are  somewhat in between the linear and
square-root dependences~\cite{17}. In the case of an intrinsic
graphene,  the Fermi energy in (10) should be replaced by the
thermal energy $kT$. Since according to (10), the density of
charged defects can be smaller than that of carriers, the
electron-hole scattering mechanism can really dominate even at low
temperatures.

In conclusion, we have asserted that an intrinsic graphene
conductivity (when both electron and hole densities exactly equal
to each other) is defined by electron-hole scattering. It has a
universal value independent of temperature. We have given an
explicit derivation based on scaling theory. This value could be
manipulated by varying permittivity of surrounding dielectrics.
When there is even a small deviation in electron and hole
densities caused by applied gate voltage the situation becomes
utterly different. In the case the conductivity is determined 
by both strong electron-hole scattering and weak external 
scattering: on defects or phonons. If phonon scattering prevails
the conductivity depends on temperature. 
We also suggest that density of charged
defects (occupancy of defects) depends on Fermi energy to explain
a sub-linear dependence of conductivity on gate voltage observed
in experiments.

The authors thank F. T. Vasko for valuable information.

\bibliography{preprint_Vyurkov}

\end{document}